\documentclass{icrc2009}

\usepackage{graphicx}   % for including figures
\usepackage[font=footnotesize]{subfig} % subfig.sty for a double column floating figure using two subfigures
\usepackage{fixltx2e}
\usepackage{url}

\newcommand{\shorttitle}[1]%
{\markboth{Proceedings of the 31\MakeLowercase{$^{st}$} ICRC, {\L}\'{o}d\'{z} 2009}{#1} }
\newcommand{\etal}{\MakeLowercase{\textit{et al. }}} % "et al."

\begin{document}
\title{VERITAS Observations of LS~I~+61${^\circ}$303 in the Fermi Era}

\author{\IEEEauthorblockN{Jamie Holder\IEEEauthorrefmark{1} 
    for the VERITAS Collaboration\IEEEauthorrefmark{2}}
                            \\
\IEEEauthorblockA{\IEEEauthorrefmark{1}Department of Physics and Astronomy and the Bartol Research Institute, University of Delaware, USA.}  
\IEEEauthorblockA{\IEEEauthorrefmark{2}see R.A. Ong \etal (these proceedings) or \url{http://veritas.sao.arizona.edu/conferences/authors?icrc2009}}}

% please write the preseter's name and short title (3-4 words maximum)
%    which will appear at the header of the even pages.
\shorttitle{J. Holder \etal VERITAS Observations of LS~I~$+61{^\circ}303$}
\maketitle

\begin{abstract}

The high-mass X-ray binary system LS~I~+61$^{\circ}$303 is well known
as a rare example of a variable Galactic GeV and TeV gamma-ray
emitter. Despite years of study, many aspects of the system remain
unclear; the nature of the compact object, the particle acceleration
mechanisms and the gamma-ray emission and absorption processes can all
be modelled in a variety of different scenarios. Here we report on a
deep exposure of LS~I~+61$^{\circ}$303made with the VERITAS array
during the 2008-2009 observing season. These are the first TeV
observations made with contemporaneous coverage at lower energies by
the LAT onboard Fermi, and as such provide a new set of constraints
for system models.
  \end{abstract}

 
\section{Introduction}

LS~I~+61$^{\circ}$~303 is a high-mass X-ray binary system at an
estimated distance of $\sim 2~\mathrm{kpc}$, composed of a compact
object and a B0 Ve star of 12.5 $M_{\odot}$ with a circumstellar
disc. The observed radio through optical emission is modulated with a
period ($P=26.4960\pm0.0028~\mathrm{days}$), which is believed to be associated
with the orbital period of the binary system \cite{gregory02,
casares05}.  New spectroscopic optical measurements by Aragona et
al. \cite{aragona09} provide a reassessment of the orbital elements,
with periastron now believed to take place at phase $\phi=0.275$ and
the orbital eccentricity $e=0.537\pm0.034$.

X-ray modulation at the orbital period has been measured with the
Rossi X-Ray Timing Explorer (RXTE) All-Sky Monitor \cite{leahy01,
wen06}, although regular observations taken every 2 days over a period
covering 6 orbital cycles in 2007-2008 with the RXTE Proportional
Counter Array (PCA) show no evidence for phase-dependent flux
modulation \cite{smith_LSI}. These same measurements, as well as later
measurements by Ray \& Hartman \cite{rayatel} reveal bright X-ray
flaring episodes lasting longer than 10 minutes, with substructure on
the timescale of a few seconds, and possible quasi-periodic
oscillations. A much shorter timescale flare, lasting 0.23 seconds,
has also been detected with the Swift Burst Alert Telescope (BAT)
\cite{dubusatel}. In both cases, the X-ray variability may be related
to other sources in the same field-of-view (e.g. \cite{munozatel}).

A gamma-ray source coincident with LS~I~+61${^\circ}$303 was first detected by COS-B
\cite{swanenburg81}; observations with the Energetic Gamma Ray
Experiment Telescope (EGRET) revealed variability on both short
($\sim$daily) and long ($\sim$monthly) timescales
\cite{tavani98}. Firm identification of the gamma-ray source with LS~I~+61${^\circ}$303
was achieved when ground-based TeV gamma-ray observatories provided an
accurate position for the site of the variable gamma-ray emission
\cite{albert_LSIa, acciari_LSIa}. 

With the exception of the Crab Nebula, LS~I~+61${^\circ}$303 is now
possibly the most heavily observed TeV source, with deep exposures by
Whipple \cite{smith07}, MAGIC \cite{albert_LSIa, albert_LSIb, albert_LSIc} and
VERITAS \cite{acciari_LSIa, acciari_LSIb} accumulated over many
years. TeV emission is only strongly detected around apastron (around phases
$\phi=0.5-0.8$), although weak evidence for low-flux emission at other
phases has been reported \cite{albert_LSIc, acciari_LSIb}. Analysis of
the complete MAGIC dataset has revealed modulation of the gamma-ray
flux close to the binary orbital period. The relatively low source
flux combined with an orbital period closely matched to the lunar
cycle means that the variability of the TeV emission with phase is
often investigated using data taken over many orbital cycles, thus
implicitly assuming no major orbit-to-orbit variability. Comparison
with observations at other wavelengths also often assumes a consistent
orbital dependence; clearly, strictly contemporaneous observations
allow more rigorous conclusions to be drawn.

The launch of the Large Area Telescope (LAT) onboard the Fermi
Gamma-ray Space Telescope provides the opportunity for such
studies. First results from LAT observations \cite{dubois09} show that
LS~I~+61${^\circ}$303 is detected continuously over all phases and that the lightcurve
shows orbit-to-orbit variability. Unlike the behaviour at TeV
energies, the emission measured by the LAT peaks at apastron. To
coincide with the first LAT observations, VERITAS conducted an
intensive observing campaign during one orbital cycle in
October/November 2008, covering phases $\phi=0.03-0.75$. Further
observations were made to overlap with a Suzaku X-ray exposure in
January/February 2009. The results of the VERITAS observations are
presented here.

\section{Observations}
VERITAS \cite{weekes02, holder09} is an array of four imaging
atmosperic Cherenkov telescopes located at the basecamp of the Fred
Lawrence Whipple Observatory near Tucson, Arizona. The array has been
fully operational since mid-2007 and has sensitivity sufficient to
detect a source with 1\% of the steady Crab Nebula flux in
$<50$~hours. Observations cover the energy range from 100~GeV to
beyond $30$~TeV with an energy resolution of 15-20\% above 300~GeV, and an
angular resolution per gamma-ray photon of $0.1^{\circ}$ at 1~TeV.

Observations of LS~I~+61${^\circ}$303 were made during clear nights with all four
telescopes operating and no major hardware problems. The source was
offset from the centre of the field-of-view by $0.5^{\circ}$ to allow
simultaneous background estimation (\textit{wobble} mode
\cite{fomin94}). In October/November 2008, 28.6 hours of data were
collected over 19 nights, from $\phi=0.03-0.75$; although we note that
observations on the last 3 nights ($\phi=0.67-0.75$) were only
possible at decreasing source elevation angles (and so lower
sensitivity). Observing conditions were excellent; this dataset is the
most sensitive TeV exposure yet obtained within a single orbital cycle
of LS~I~+61${^\circ}$303. Observations in January/February 2009 were planned to coincide
with a 130~ks Suzaku exposure beginning 2009-01-25 at 16:22
UTC. 8.2~hours of observations were made on 5 nights, from
$\phi=0.64-0.91$. Poor weather resulted in a four-day gap in coverage,
from $\phi=0.68-0.83$. All of these observations are summarized in
Table~\ref{observations}.

\section{Results}
The data were analysed using standard VERITAS analysis tools
\cite{acciari_LSIa} and with gamma-ray selection cuts suitable for a
point-like source with a spectrum similar to that of the Crab
Nebula. The background in the source region was estimated using the
``ring-background'' method \cite{berge07}. The mean elevation at which
these observations were made is $57^{\circ}$, resulting in an energy
threshold (defined as the peak of the differential gamma-ray rate for
a Crab Nebula-like spectrum) for this analysis of 440~GeV. All results
have been verified using two independent analysis chains.

\begin{figure*}[th]
  \centering
  \includegraphics[width=5in]{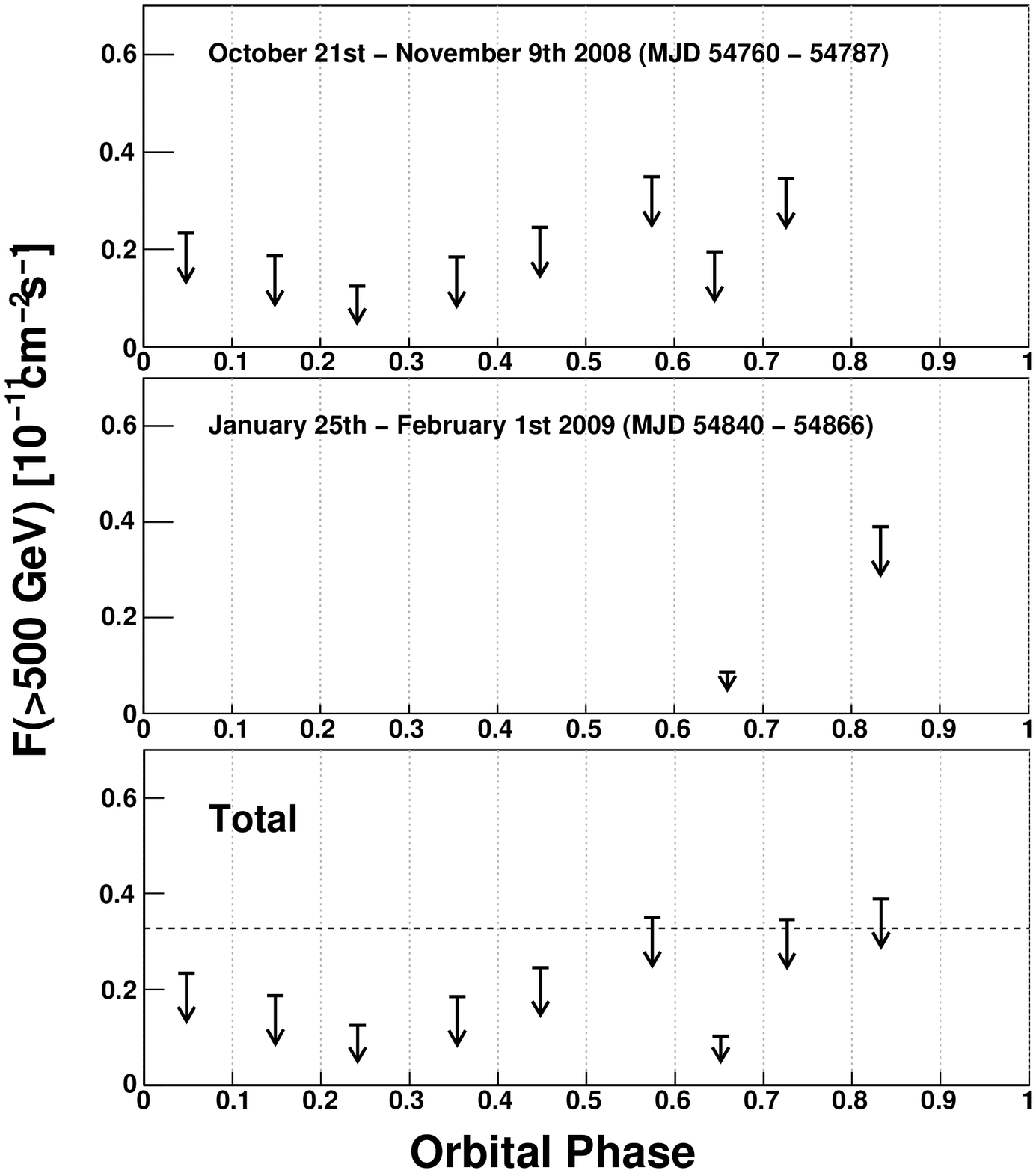}
  \caption{Integral upper flux limits to the emission from LS~I~+61${^\circ}$303 above
    500~GeV as a function of the orbital phase. Limits are at the 95\%
    confidence level assuming a power law spectral index
    $\alpha=-2.5$. Arrows are placed at the mean phase of observations
    within each of the 10 phase bins. The dashed line indicates 5\% of
    the integral flux from the Crab Nebula above the same threshold }
  \label{lightcurve}
 \end{figure*}

The complete 36.8~hour exposure results in a $3.4\sigma$ excess and a
mean integral flux above 500~GeV of
($6.2\pm2.6$)$\times10^{-13}$ph~cm$^{-2}$s$^{-1}$. Table~\ref{observations}
shows the results for each of the individual night's observations. The
quoted significance values are not corrected for statistical trials
(24 nights = 24 trials). We find no significant evidence for emission
on any single night, pre- or post-trials. Figure~\ref{lightcurve} shows flux
upper limits as a function of orbital phase for the two orbits covered
here, and for both orbits combined.

\section{Discussion}
The lack of strong emission from LS~I~+61${^\circ}$303 during these observations is
somewhat suprising, but does not necessarily contradict previous
measurements. The conclusion of previously published MAGIC and VERITAS
results is that the source is, on average, detected at TeV energies in
the phase range $\phi=0.5-0.8$. Low statistics have not allowed us to
draw firm conclusions about the light curve shape in this region, or
to address the possibility of orbit-to-orbit variability. As noted
above, the apastron coverage during this campaign was limited, with
large data gaps. For example, the coverage between phases 0.72 and
0.82 consists of only 40 minutes at $47^{\circ}$ elevation; the peak
of the TeV emission might well have occured between these phases for
these orbits and been undetectable.

Much stronger conclusions can be drawn for the October/November 2008
observations over phase ranges $\phi=0.0-0.65$, with a large dataset
of nightly observations at high elevation angles. The source is not
detected by VERITAS during any of the phase bins in this range, while
strictly contemporaneous Fermi measurements show the GeV lightcurve
peaking during periastron (around $\phi=0.275$)
\cite{dubois09}. Clearly, the GeV and TeV flux states are not
correlated. 

There are now many models in the literature which describe high energy
processes in LS~I~+61${^\circ}$303 (e.g. \cite{bosch06, bednarek06,
gupta06, dubus06a, dubus06b, romero07, zdziarski08,
sierpowska09}. These consider both accretion-driven and pulsar-wind
scenarios and invoke both hadronic and leptonic relativistic particle
populations. Their predictions are often fit to mean broadband
spectral energy distributions (SEDs) gathered from observations made
at different times in different wavelengths. The evidence presented
here for a lack of correlation between the TeV and GeV fluxes shows
that the orbital dependence of the emission in these wavebands is not
the same, providing an additional constraint for the models to
address. A number of processes could be important in this; the energy
dependence of photon-photon absorption leads to differing opacities
for GeV and TeV emission around the
orbit(e.g. \cite{sierpowska09}). Synchrotron loss rates and inverse
Compton efficiencies are both also expected to vary as the compact
object encounters changing stellar wind densities and photon fields
\cite{dubus08}, which will modify the TeV and GeV fluxes differently.

% see \section{Examples of \LaTeX\  instructions}  \subsection{Tables}
  \begin{table*}[th]
  \caption{Daily results for VERITAS observations of LS~I~+61${^\circ}$303 in 2008-2009}
  \label{observations}
  \centering
  \begin{tabular}{|c|c|c|c|c|c|c|c|c|c|}
  \hline
   Date  &  MJD & Orbital Phase & Exposure(mins) & Mean Elevation & ON & OFF & alpha & significance \\
   \hline 
    20081021 & 54760 & 0.03 & 120 &  59$^{\circ}$ &  45 & 421 & 0.11 & -0.1 \\
    20081022 & 54761 & 0.07 & 100 &  57$^{\circ}$ &  36 & 245 & 0.11 &  1.6 \\
    20081023 & 54762 & 0.11 & 100 &  59$^{\circ}$ &  35 & 212 & 0.11 &  2.0 \\
    20081024 & 54763 & 0.15 & 20  &  55$^{\circ}$ &  2  & 42  & 0.11 & -1.3 \\
    20081025 & 54764 & 0.18 & 124 &  56$^{\circ}$ &  25 & 269 & 0.11 & -0.9 \\
    20081026 & 54765 & 0.22 & 140 &  57$^{\circ}$ &  31 & 282 & 0.11 &  0.0 \\
    20081027 & 54766 & 0.25 & 100 &  58$^{\circ}$ &  21 & 218 & 0.11 & -0.6 \\
    20081028 & 54767 & 0.29 & 40  &  59$^{\circ}$ &  18 & 93  & 0.11 &  2.1 \\
    20081029 & 54768 & 0.33 & 100 &  58$^{\circ}$ &  29 & 183 & 0.11 &  1.8 \\
    20081030 & 54769 & 0.37 & 125 &  59$^{\circ}$ &  24 & 210 & 0.11 &  0.2 \\
    20081031 & 54770 & 0.41 & 60  &  58$^{\circ}$ &  17 & 87  & 0.11 &  2.0 \\
    20081101 & 54771 & 0.44 & 80  &  60$^{\circ}$ &  21 & 158 & 0.11 &  0.8 \\
    20081102 & 54772 & 0.48 & 80  &  59$^{\circ}$ &  28 & 155 & 0.11 &  2.4 \\
    20081104 & 54774 & 0.56 & 100 &  60$^{\circ}$ &  32 & 188 & 0.11 &  2.1 \\
    20081105 & 54775 & 0.60 & 68  &  57$^{\circ}$ &  17 & 113 & 0.11 & -1.2 \\
    20081106 & 54776 & 0.63 & 180 &  58$^{\circ}$ &  33 & 330 & 0.11 &  0.5 \\
    20081107 & 54777 & 0.67 & 80  &  53$^{\circ}$ &  25 & 141 & 0.11 & -2.1 \\
    20081108 & 54778 & 0.71 & 60  &  53$^{\circ}$ &  8  & 92  & 0.11 &  0.6 \\
    20081109 & 54779 & 0.75 & 40  &  47$^{\circ}$ &  9  & 61  & 0.11 &  0.8 \\
   \hline 			          		       
    20090125 & 54856 & 0.64 & 115 &  57$^{\circ}$ &  25 & 219 & 0.11 &  0.2 \\
    20090126 & 54857 & 0.68 & 80  &  58$^{\circ}$ &  8  & 123 & 0.11 & -1.6 \\
    20090130 & 54861 & 0.83 & 100 &  56$^{\circ}$ &  14 & 69  & 0.11 &  1.9 \\
    20090131 & 54862 & 0.87 & 114 &  55$^{\circ}$ &  21 & 195 & 0.11 & -0.1 \\
    20090201 & 54863 & 0.91 & 80  &  54$^{\circ}$ &  23 & 151 & 0.11 &  1.4 \\
  \hline
  \end{tabular}
  \end{table*}

% see \section{Examples of \LaTeX\  instructions} and \subsection{Figures}
% An example of a double column floating figure using two subfigures.
% The double column figure must be placed in the source text file 
%         within the text of the previous page

\section{Acknowledgements}

This research is supported by grants from the US Department of
Energy, the US National Science Foundation, and the Smithsonian
Institution, by NSERC in Canada, by Science Foundation Ireland, and
by STFC in the UK. We acknowledge the excellent work of the technical
support staff at the FLWO and the collaborating institutions in the
construction and operation of the instrument.

\end{document}